\documentclass{ifacconf}

\usepackage{graphicx}      
\usepackage{natbib}        

\usepackage{amssymb}          
\usepackage{amsmath}          
\usepackage{graphicx}         
\usepackage{enumerate}
\usepackage{color}
\usepackage{algorithm}
\usepackage{algorithmicx}

\begin{document}
\begin{frontmatter}

\title{Direct identification of fault estimation filter for sensor faults \thanksref{footnoteinfo}}

\thanks[footnoteinfo]{The research leading to these results has received funding from the European Union's Seventh Framework Programme (FP7-RECONFIGURE/2007--2013) under grant agreement No. 314544.}

\author[first]{Yiming Wan}
\author[first]{Tamas Keviczky}
\author[first]{Michel Verhaegen}

\address[first]{Delft University of Technology,
   2628CD, Delft, The Netherlands (e-mail: y.wan@tudelft.nl; t.keviczky@tudelft.nl; m.verhaegen@tudelft.nl).}

\begin{abstract}                
We propose a systematic method to directly identify a sensor fault estimation filter from plant input/output data collected under fault-free condition.
This problem is challenging, especially when omitting the step of building an explicit state-space plant model in data-driven design, because the inverse of the underlying plant dynamics is required and needs to be stable. 
We show that it is possible to address this problem by relying on a system-inversion-based fault estimation filter that is parameterized using identified Markov parameters.
Our novel data-driven approach improves estimation performance by avoiding the propagation of model reduction errors originating from identification of the state-space plant model into the designed filter.
Furthermore, it allows additional design freedom to stabilize the obtained filter under the same stabilizability condition as the existing model-based system inversion.
This crucial property enables its application to sensor faults in unstable plants, where
existing data-driven filter designs could not be applied so far due to the lack of such stability guarantees (even after stabilizing the closed-loop system).
A numerical simulation example of sensor faults in an unstable aircraft system illustrates the effectiveness of the proposed new method.
\end{abstract}


\end{frontmatter}

\section{Introduction}
Model-based fault diagnosis techniques for linear dynamic systems have been well established during the past two decades \citep{ChenPatton1999, Ding2013book}. However, an explicit and accurate system model is often unknown in practice. In such situations, a conventional approach first identifies the plant model from system input/output (I/O) data, and then adopts various model-based fault diagnosis methods \citep{Simani2003, Patward2005, Manuja2009}. Without explicitly identifying a plant model, recent research efforts investigate data-driven approaches to directly construct a fault diagnosis system for additive sensor or actuator faults by utilizing the link between system identification and the model-based fault diagnosis methods \citep{Qin2012, Ding2014book}. These recent direct data-driven approaches simplify the design procedure by skipping the realization of an explicit plant model, while at the same time allow developing systematic methods to address the same fault diagnosis performance criteria as the existing model-based approaches.

Existing methods for data-driven fault detection and isolation construct residual generators with either the parity vector/matrix \citep{Ding2009JPC} or the Markov parameters (or impulse response parameters) \citep{Dong2012a, Dong2012b} that can be identified from data. Compared to generating residual signals sensitive to faults, fault estimation is much more involved.
\cite{DuniaQin1998} and \cite{Qin2012} proposed to reconstruct faults by minimizing the squared reconstructed prediction error in the residual subspace of a latent variable model. However, fault reconstructability of this approach was limited by the dimension of the residual subspace, especially when using a dynamic latent variable model \citep{DuniaQin1998}.
Chapter 10 of \cite{Ding2014book} first constructed a diagnostic observer realized with the identified parity vector/matrix, and then addressed faults as augmented state variables. This augmented observer scheme, however, imposed certain limitations on how fault signals vary with time.
In contrast, \cite{Dong2012c} constructed a system-inversion-based fault estimator using the identified Markov parameters, without any assumptions on the dynamics of fault signals. The drawback of this system-inversion-based method is that it cannot be applied to sensor faults in an unstable open-loop plant because its underlying system inverse used for the data-driven design is unstable in this case. In order to address this above drawback, we have recently proposed a receding horizon fault estimator by following a least-squares (LS) formulation of the fault estimation problem, and developed an $H_2 / H_\infty$ optimal design to compensate for identification errors of the Markov parameters \citep{WanKevicVerh2014}.
This receding horizon method processes a batch of measurements at each time instant, thus may require increased computational effort.

This paper focuses on the direct data-driven design problem for a sensor fault estimation filter. This problem is challenging, especially when omitting the step of building an explicit state-space plant model in data-driven design, because the inverse of the underlying plant dynamics is required and needs to be stable. In order to pave the way for the data-driven design, we first construct a system-inversion-based fault estimation filter based on the dynamics of the one-step-ahead predicted residual signal. The system inverse is divided into two parts: the open-loop left inverse, and the feedback from the residual reconstruction error to stabilize the inverse dynamics. This turns out to be stabilizable as long as the subsystem from faults to the outputs has no unstable invariant zeros. Our data-driven design method is obtained by parameterizing the above two parts of the inverse dynamics with the predictor Markov parameters identified from data.

Compared to the model-based approach based on an identified plant model, our direct data-driven design improves estimation performance by avoiding the propagation of model reduction errors originating from identification of the state-space plant model into the designed filter.
Moreover, our proposed new method allows additional design freedom to stabilize the obtained filter under the same stabilizability condition as the existing model-based system inversion. This important additional property enables its application to sensor faults in unstable plants, where
existing data-driven filter designs \citep{Dong2012c} could not be applied so far due to the lack of such stability guarantees (even after stabilizing the closed-loop system).
We also analyze the relationship between our novel data-driven design described in this paper, and our recently proposed moving horizon fault estimation method in \cite{WanKevicVerh2014}.
The presented new data-driven filter achieves better computational efficiency at the cost of minor performance loss and a more strict condition required for unbiasedness.
The above significant advances to the state-of-the-art in data-driven fault estimation are illustrated via a numerical simulation example of an unstable aircraft system.

\section{Preliminaries and problem formulation}
\subsection{Notations}
For a matrix $X$, $X^{-}$ represents the left inverse satisfying $X^{-} X = I$, and $X^{(1)}$ represents the generalized inverse satisfying
\begin{equation}\label{eq:ginv}
X X^{(1)} X = X.
\end{equation}
The $i^{\mathrm{th}}$ column of $X$ is denoted by $X^{[i]}$.
For the state-space model $\left( A, B, C, D \right)$ or the sequence of Markov parameters $H_0, H_1, \cdots, H_{L-1}$,
let $\mathcal{O}_L$ and $\mathcal{T}_L$ denote the extended observability matrix with $L$ block elements and the lower triangular block-Toeplitz matrix with $L$ block columns and rows, respectively,  i.e.,
\begin{align}
& \mathcal{O}_{L} \left( A, C \right) = \left[ \begin{smallmatrix}
                          C \\
                          C A \\
                          \vdots \\
                          C A^{L-1}
                        \end{smallmatrix} \right],
                        \mathcal{T}_{L} \left( \{ H_i \} \right) = \left[ \begin{smallmatrix}
                          H_0 & 0 & \ldots & 0 \\
                          H_1 & H_0 & \ddots & \vdots \\
                          \vdots & \vdots & \ddots & 0  \\
                          H_{L-1} & H_{L-2} & \cdots & H_0
                        \end{smallmatrix} \right],
                        \label{eq:OL_TLmarkov} \\
& \text{or}\;\; \mathcal{T}_{L} \left( A, B, C, D \right) = \left[ \begin{smallmatrix}
                          D & 0 & \ldots & 0 \\
                          C B & D & \ddots & \vdots \\
                          \vdots & \vdots & \ddots & 0  \\
                          C A^{L -2} B & C A^{L -3} B & \cdots & D
                        \end{smallmatrix} \right].
                        \label{eq:TLu_ss}
\end{align}
$\mathbb{E}$ represents the mathematical expectation.

\subsection{System description}
We consider linear discrete-time systems governed by the following state-space model:
\begin{equation}\label{eq:sys}
\begin{aligned}
\xi (k+1) &= A \xi(k) + Bu(k) + E f(k) + F w(k) \\
y(k) &= C \xi (k) + Du(k) + G f(k) + v(k).
\end{aligned}
\end{equation}
Here $\xi (k)\in\mathbb{R}^{n}$, $y(k)\in\mathbb{R}^{n_y}$, and $u(k)\in\mathbb{R}^{n_u}$ represent the state, the output measurement, and the known control input at time instant $k$, respectively. The process noise $w(k)\in \mathbb{R}^{n_w}$ and the measurement
noise $v(k) \in \mathbb{R}^{n_v}$ are white zero-mean Gaussian, with covariance matrices
$\mathrm{E}\left( w(k) w^\mathrm{T}(k) \right) = Q$,
$\mathrm{E}\left( v(k) v^\mathrm{T}(k) \right) = R$,
$\mathrm{E}\left( w(k) v^\mathrm{T}(k) \right) = 0$.
$f(k)\in\mathbb{R}^{n_f}$ is the unknown fault signal to be estimated.
$A, B, C, D, E, F, G$ are constant real matrices, with bounded norms and appropriate dimensions.

The following assumption is standard in Kalman filtering \citep{Kailath2000} and subspace identification \citep{Chiuso2007, Chiuso2007a}:
\begin{assum}\label{ass:detect_control}
The pair $\left( A, C \right)$ is assumed detectable; and there are no uncontrollable modes of $( A, F Q^{\frac{1}{2}} )$ on the unite circle, where $Q^{\frac{1}{2}} \cdot ( Q^{\frac{1}{2}} )^\mathrm{T} = Q$ is the covariance matrix of $w(k)$.
\end{assum}

We consider additive sensor faults in this paper, i.e.,
\begin{equation}\label{eq:fault_model}
E = 0_{n_x \times 1},\;\; G = I^{[j]}
\end{equation}
for faults of the $j^{\rm{th}}$ sensor,
with $X^{[j]}$ representing the $j^{\mathrm{th}}$ column of a matrix $X$. As in \cite{Dong2012c}, we adopt the following common assumption for sensor faults:

\begin{assum}\label{ass:fault_rank}
$\mathrm{rank}\left(G\right) = n_f$.
\end{assum}

For data-driven design without knowing the system matrices in (\ref{eq:sys}), it should be noted that in practice data collected under faulty conditions may be seldomly available, or if recorded then without a reliable fault description \citep{Ding2014JPC}. Hence we make the assumption as below:
\begin{assum}\label{ass:data}
Only I/O data collected under the fault-free condition are used in our data-driven design.
\end{assum}

No assumption is made in this paper about how the fault signals $f(k)$ evolve with time.

\subsection{Problem formulation}
With Assumption \ref{ass:detect_control}, the system (\ref{eq:sys}) admits the innovation form given by
\begin{equation}\label{eq:innovationform}
\begin{aligned}
x(k+1) &= A x(k) + B u(k) + E f(k) + K e(k) \\
y(k) &= C x(k) + D u(k) + G f(k) + e(k).
\end{aligned}
\end{equation}
where $K$ is the steady-state Kalman gain, $\left\{e(k)\right\}$ is the zero-mean innovation process with the covariance matrix $\Sigma_e$. Then $e(k)$ can be eliminated from the first equation of (\ref{eq:innovationform}) to yield the one-step-ahead predictor form
\begin{equation}\label{eq:predictor}
\begin{aligned}
x(k+1) &= \Phi x(k) + \tilde B u(k) + \tilde E f(k) + K y(k) \\
y(k) &= C x(k) + D u(k) + G f(k) + e(k),
\end{aligned}
\end{equation}
with $\Phi = A - KC$, $\tilde B = B - KD$, and $\tilde E = E - KG$.
The sensor fault direction matrices $\tilde E$ and $G$ in the predictor form (\ref{eq:predictor}) can be explicitly written as \begin{equation}\label{eq:fault_model_predictor}
\tilde E = - K^{[j]},\;\; G = I^{[j]}
\end{equation}
for faults of the $j^{\rm{th}}$ sensor according to (\ref{eq:fault_model}).

%

Denote the predictor Markov parameters by
\begin{equation}\label{eq:markov_param}
\begin{aligned}
& H_i^u = \left\{ \begin{array}{ll}
                  D & i=0 \\
                  C \Phi^{i-1} \tilde B & i>0
                \end{array} \right. , \;
H_i^y = \left\{ \begin{array}{ll}
                  0 & i=0 \\
                  C \Phi^{i-1} K & i>0
                \end{array} \right. , \\
& H_i^f = \left\{ \begin{array}{ll}
                  G & i=0 \\
                  C \Phi^{i-1} \tilde E & i>0
                \end{array} \right. .
\end{aligned}
\end{equation}

With access only to the closed-loop data collected under the fault-free condition,
the conventional model-based approach needs to identify the state-space model (\ref{eq:predictor}) for the fault estimation filter design. Such an identification algorithm follows three steps:
(\romannumeral1) consistent LS estimation of the sequence of Markov parameters related to the fault-free subsystem $\left( \Phi, \left[\begin{array}{cc}
 \tilde B & K \end{array} \right], C, \left[ \begin{array}{cc} D & 0 \end{array} \right] \right)$, i.e.,
\begin{equation}\label{eq:Xi_Hi}
\Xi = \left[ \begin{array}{cccccc}
               H_{p}^u & H_{p}^y & \cdots & H_{1}^u & H_{1}^y & H_{0}^u
             \end{array}
 \right];
\end{equation}
(\romannumeral2) state-space realization of the fault-free subsystem
$\left( \Phi, \left[\begin{array}{cc}
 \tilde B & K \end{array} \right], C, \left[ \begin{array}{cc} D & 0 \end{array} \right] \right)$;
(\romannumeral3) construction of the fault direction matrices $\tilde E$ and $G$ according to (\ref{eq:fault_model_predictor}). The first two identification steps above can follow the predictor-based subspace identification (PBSID) method in \cite{Chiuso2007, Chiuso2007a}. With the identified state-space model, existing model-based design approaches can be adopted. A disadvantage of the above design procedure is that the model reduction errors introduced in the state-space realization step would propagate into the fault estimation filter and might result in large fault estimation errors.

In order to avoid propagating the above model reduction errors into the designed filter, this paper aims to directly construct a stable sensor fault estimation filter with the Markov parameters $\Xi$ identified from data.

\section{System-inversion-based fault estimation filter using predictor form}
\label{sect:sspred_fest}
As the foundation for our data-driven design, we construct a system-inversion-based fault estimation filter in this section using the state-space model of the predictor (\ref{eq:predictor}).

\subsection{Open/Closed-loop left inverse}
From (\ref{eq:predictor}), we construct a residual generator as follows:
\begin{equation}\label{eq:res_gen_ss}
\begin{aligned}
\hat x(k+1) &= \Phi \hat x(k) + \tilde B u(k) + K y(k) \\
r(k) &= y(k) - C \hat x(k) - D u(k),
\end{aligned}
\end{equation}
whose residual dynamics is
\begin{subequations}\label{eq:res_dyn_ss}
\begin{align}
\tilde x(k+1) &= \Phi \tilde x(k) + \tilde E f(k) \label{eq:res_dyn_1} \\
r(k) &= C \tilde x(k) + G f(k) + e(k), \label{eq:res_dyn_2}
\end{align}
\end{subequations}
with $\tilde x(k) = x(k) - \hat x(k)$.

By multiplying (\ref{eq:res_dyn_2}) with $G^-$, it follows that $f(k)$ can be reconstructed as
\begin{equation*}
f(k) = G^- \left( r(k) - C \tilde x(k) - e(k) \right).
\end{equation*}
Substituting the above equation into (\ref{eq:res_dyn_1}) then yields the following left inverted system:
\begin{subequations}\label{eq:open_loop_inv}
\begin{align}
\tilde x (k+1) &= \Phi_1 \tilde x (k) + B_1 \left( r(k) - e(k)  \right) \label{eq:open_loop_inv_dyn}  \\
f(k) &= C_1 \tilde x (k) + D_1 \left( r(k) - e(k)  \right) \label{eq:open_loop_inv_out}
\end{align}
\end{subequations}
with
\begin{gather}
\Phi_1 = \Phi - \tilde E G^- C, \;
B_1 = \tilde E G^-, \label{eq:Phi1_B1} \\
C_1 = - G^- C, \;
D_1 = G^-. \label{eq:C1_D1}
\end{gather}
Since the innovation signal $e(k)$ and the initial state $\tilde x(0)$ are unknown, we construct the following system based on the inverted system (\ref{eq:open_loop_inv}) by ignoring $e(k)$ and replacing $\tilde x(k)$ and $f(k)$ with the state estimate $x_r (k)$ and the fault estimate $\hat f(k)$:
\begin{subequations}\label{eq:open_loop_fest}
\begin{align}
x_r (k+1) &= \Phi_1 x_r (k) + B_1 r(k) \label{eq:open_loop_fest_dyn}  \\
\hat f(k) &= C_1 x_r (k) + D_1 r(k) \label{eq:open_loop_fest_out}.
\end{align}
\end{subequations}
It is desired that the left inverse (\ref{eq:open_loop_fest}) is stable such that, starting from any arbitrary estimate $x_r(0)$ of the initial state $\tilde x(0)$, unbiasedness of the estimates $x_r(k)$ and $\hat f(k)$ can be achieved asymptotically. However, it is not guaranteed that the left inverse (\ref{eq:open_loop_fest}) is stable.

\begin{figure}
\begin{center}
\includegraphics[width=7cm]{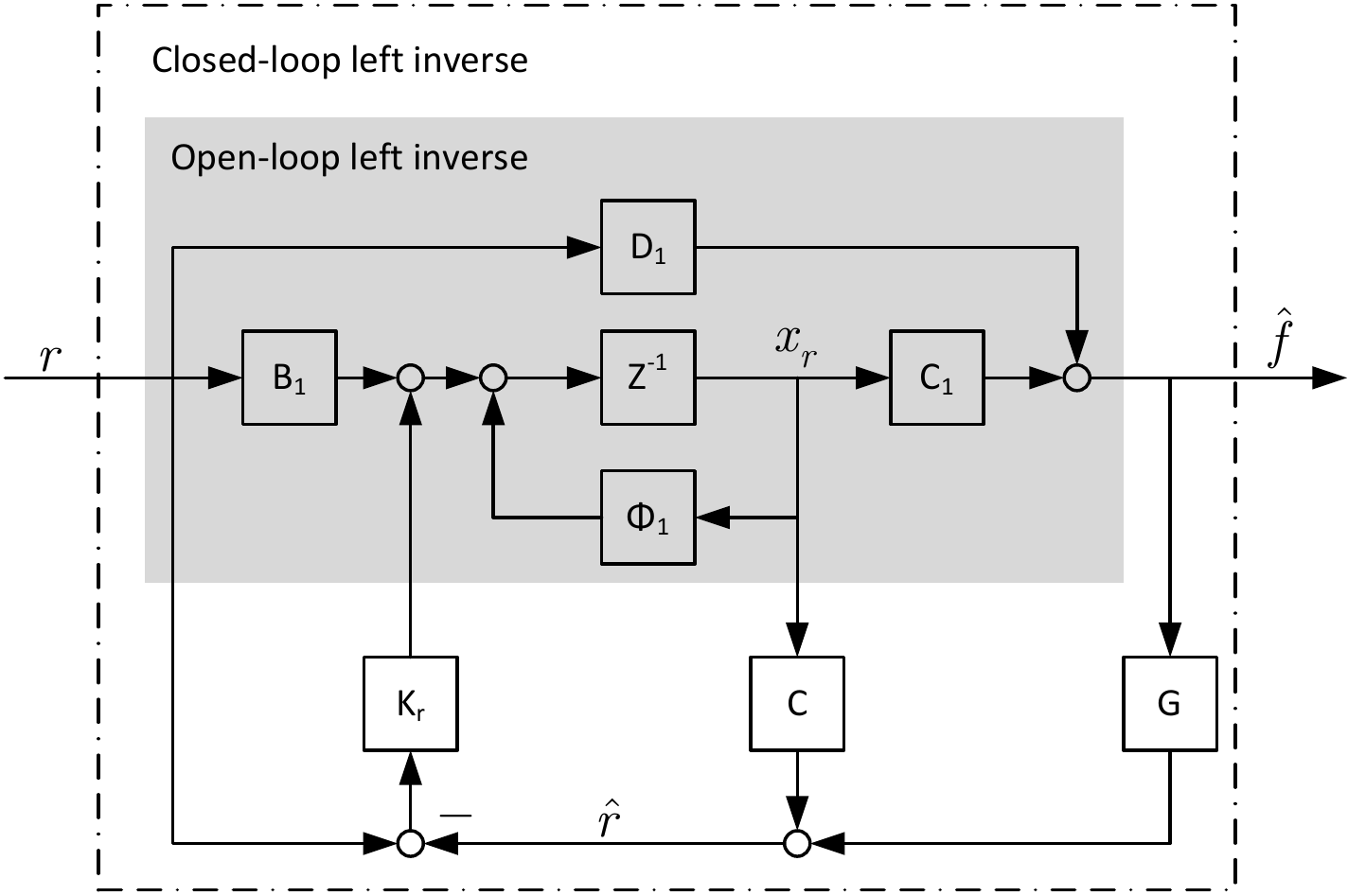}    
\caption{Structure of the closed-loop left inverse}
\label{fig:sysinv_structure}
\end{center}
\end{figure}

Next, we stabilize the inverted system (\ref{eq:open_loop_fest}) by feeding the residual reconstruction error back into (\ref{eq:open_loop_fest_dyn}). Based on the state estimate $x_r(k)$ and the fault estimate $\hat f(k)$ in (\ref{eq:open_loop_fest_out}), the residual signal $r(k)$ can be reconstructed as
\begin{align}\label{eq:res_reconstruct}
\hat r(k) &= C x_r(k) + G \left( C_1 x_r (k) + D_1 r(k) \right) \nonumber \\
&= C_2 x_r (k) + D_2 r(k)
\end{align}
according to (\ref{eq:res_dyn_2}), with
\begin{gather}\label{eq:C2_D2}
C_2 = \left( I - G G^- \right) C, \;
D_2 = G G^-.
\end{gather}
Then we construct the following closed-loop left inverse by feeding the residual reconstruction error $r(k) - {\hat r}(k)$ back into the open-loop left inverse (\ref{eq:open_loop_fest}):
\begin{equation}\label{eq:sysinv_closed_loop}
\begin{aligned}
x_r (k+1) =& \Phi_1 x_r (k) + B_1 r(k) + K_r \left( r(k) - \hat r(k) \right) \\
 =& \Phi_2 x_r(k) + B_2 r(k) \\
\hat f(k) =& C_1 x_r (k) + D_1 r(k)
\end{aligned}
\end{equation}
with
\begin{gather}\label{eq:Phi2_B2}
\Phi_2 = \Phi_1 - K_r C_2, \;
B_2 = B_1 + K_r \left( I - D_2 \right).
\end{gather}
Here ``open-loop'' and ``closed-loop'' refer to the absence/presence of the feedback from the residual reconstruction error in the two inverted systems (\ref{eq:open_loop_fest}) and (\ref{eq:sysinv_closed_loop}).
The structure of the closed-loop left inverse is illustrated in Figure \ref{fig:sysinv_structure}.

It is worth noting that the data-driven filter design in \cite{Dong2012c} considered only the open-loop left inverse whose stability is not guaranteed.
It should also be pointed out that similarly to the simultaneous state and input estimation filter proposed in \cite{Gill2007}, the closed-loop left inverse (\ref{eq:sysinv_closed_loop}) produces both the state estimate $x_r(k)$ and the fault estimate $\hat f(k)$. However, the proposed closed-loop inverse (\ref{eq:sysinv_closed_loop}) has a more structured formulation, i.e., the combination of the open-loop left inverse and the feedback from the residual reconstruction error. Such a formulation enables our data-driven design in Section \ref{sect:design_using_Markov}.

\subsection{Stabilizability and unbiasedness}

By defining $\tilde x_r (k) = \tilde x (k) - x_r(k)$ and $\tilde f(k) = f(k) - \hat f(k)$, we can obtain the dynamics of the fault estimation error as
\begin{equation}\label{eq:fest_err_dyn}
\begin{aligned}
\tilde x_r (k+1) &= \Phi_2 \tilde x_r (k) - B_2 e(k) \\
                 &= \left( \Phi_1 - K_r C_2 \right) \tilde x_r (k) - B_2 e(k) \\
\tilde f(k) &= C_1 \tilde x_r(k) - D_1 e(k)
\end{aligned}
\end{equation}
according to (\ref{eq:res_dyn_ss}), (\ref{eq:Phi1_B1})-(\ref{eq:C1_D1}), and (\ref{eq:sysinv_closed_loop})-(\ref{eq:Phi2_B2}). Therefore, if $\left( \Phi_1, C_2 \right)$ is stabilizable, there exists a stabilizing gain $K_r$ in (\ref{eq:fest_err_dyn}), such that the obtained fault estimates are asymptotically unbiased, i.e.,
$\mathop {\lim }\limits_{k \rightarrow \infty} \mathbb{E} \left(\tilde f(k)\right) = 0$.

\begin{thm}\label{thm:stabilizability}
$( \Phi_1, C_2 )$ is stabilizable if the fault subsystem $( \Phi, \tilde E, C, G )$ has no unstable invariant zeros.
\end{thm}

The proof is given in Appendix \ref{app:proof_stabilizability}.

\subsection{Fault estimation filter}
So far we have constructed the closed-loop left inverse (\ref{eq:sysinv_closed_loop}) to estimate faults from the residual signal generated by (\ref{eq:res_gen_ss}).
By cascading the residual generator (\ref{eq:res_gen_ss}) and the closed-loop left inverse (\ref{eq:sysinv_closed_loop}), we obtain the following fault estimation filter that produces fault estimates from the system inputs and outputs:
\begin{equation}\label{eq:sfestfilter}
\begin{aligned}
\left[ \begin{matrix}
x_r (k+1) \\
\hat x (k+1)
\end{matrix} \right]
=& \left[ \begin{matrix}
\Phi_2 & -B_2 C \\
0 & \Phi
\end{matrix} \right]
\left[ \begin{matrix}
x_r (k) \\
\hat x (k)
\end{matrix} \right] +
\left[ \begin{matrix}
-B_2 D & B_2 \\
\tilde B & K
\end{matrix} \right] z(k) \\
\hat f(k) =&
\left[ \begin{matrix}
C_1 & -D_1 C
\end{matrix} \right]
\left[ \begin{matrix}
x_r (k) \\
\hat x (k)
\end{matrix} \right]
+ \left[ \begin{matrix}
-D_1 D & D_1
\end{matrix} \right] z(k)
\end{aligned}
\end{equation}
where
\begin{equation}\label{eq:z}
z(k) = \left[ \begin{matrix}
u^\mathrm{T}(k) & y^\mathrm{T}(k)
\end{matrix} \right]^\mathrm{T}.
\end{equation}
With $x_f(k) = x_r(k) + \hat x(k)$, the above fault estimation filter (\ref{eq:sfestfilter}) can be reduced as follows by eliminating the unobservable modes:
\begin{equation}\label{eq:sfestfilter_reduced}
\begin{aligned}
x_f (k+1) =& \left( \Phi_1 - K_r C_2 \right) x_f (k) + \left( B_f - K_r D_{f,2} \right) u(k) \\
&+ \left( K_f + K_r G_{f,2} \right) y(k) \\
\hat f (k) =& C_1 x_f(k) + D_{f,1} u(k) + D_{1} y(k),
\end{aligned}
\end{equation}
where $\Phi_1$, $C_1$, $D_1$ and $C_2$ are defined in (\ref{eq:Phi1_B1}), (\ref{eq:C1_D1}), and (\ref{eq:C2_D2}), respectively, and
\begin{subequations}\label{eq:BDKGf}
\begin{align}
B_f &= \tilde B - \tilde E G^- D, \; &D_{f,2} &= \left( I - G G^- \right) D, \label{eq:BDf}  \\
K_f &= K+ \tilde E G^-, \; &G_{f,2} &= \left( I - G G^- \right), \label{eq:KGf}  \\
D_{f,1} &= - G^- D.  & & \label{eq:Df}
\end{align}
\end{subequations}

\section{Fault estimation filter design using Markov parameters}\label{sect:design_using_Markov}
In this section, we propose our Markov-parameter based design of the fault estimation filter (\ref{eq:sfestfilter_reduced}) by exploiting its extended form over a time window.

\subsection{Extended form of the fault estimation filter}
With $k_0 = k-L+1$, define stacked data vectors in a sliding window $\left[ k_0, k \right]$ as $\mathbf{z}_{k,L}$,  $\mathbf{r}_{k,L}$, $\mathbf{f}_{k,L}$, and $\mathbf{e}_{k,L}$, respectively for the signals $z$, $r$, $f$, and $e$, e.g.,
\begin{equation}\label{eq:ukL}
\mathbf{z}_{k,L} = \left[ \begin{array}{ccc}
                                  z^\mathrm{T}\left(k_0\right) & \cdots & z^\mathrm{T}\left(k\right)
                                \end{array} \right]^\mathrm{T}.
\end{equation}
Define $\mathbf{T}_{L}^{f}$ and $\mathbf{T}_{L}^{z}$ as the lower triangular block-Toeplitz matrices, i.e.,
\begin{align}
\mathbf{T}_{L}^{f} = \mathcal{T}_L ( \Phi, \tilde E, C, G ),
\mathbf{T}_{L}^{z} = \mathcal{T}_L ( \Phi, [\begin{array}{cc} \tilde B & K \end{array}], -C, [\begin{array}{cc} -D& I \end{array}] ), \label{eq:TLz}
\end{align}
with $\mathcal{O}_L$ and $\mathcal{T}_L$ defined in (\ref{eq:OL_TLmarkov}) and (\ref{eq:TLu_ss}). According to (\ref{eq:res_gen_ss}) and (\ref{eq:res_dyn_ss}), the stacked residual signal $\mathbf{r}_{k,L}$ over the time window $\left[ k_0, k \right]$ can be written in the extended form
\begin{equation}\label{eq:resL_compute}
\begin{aligned}
\mathbf{r}_{k,L} &= \mathcal{O}_L \left( \Phi, -C \right) \cdot \hat x(k_0) + \mathbf{T}_L^z \mathbf{z}_{k,L} \\
&= \mathcal{O}_L \left( \Phi, C \right) \cdot \tilde x(k_0) + \mathbf{T}_L^f \mathbf{f}_{k,L} + \mathbf{e}_{k,L}
\end{aligned}
\end{equation}
where $\mathbf{z}_{k,L}$ is defined in (\ref{eq:ukL}), and $z(k)$ is defined in (\ref{eq:z}).

By assuming the initial state $x_r(k_0) = 0$ in the closed-loop left inverses (\ref{eq:sysinv_closed_loop}), the stacked fault estimates $\mathbf{\hat f}_{k,L}$ over the time window $\left[ k_0, k \right]$ can be written in the extended form
\begin{equation}\label{eq:festkL_inv}
\mathbf{\hat f}_{k,L} = \mathcal{K}_L \cdot \mathbf{r}_{k,L} \;\;\text{with}\;\;
\mathcal{K}_L = \mathcal{T}_L \left( \Phi_2, B_2, C_1, D_1 \right).
\end{equation}
With tedious but straightforward manipulations, the block-Toeplitz matrix $\mathcal{K}_L$ defined in (\ref{eq:festkL_inv}) can be rewritten as
\begin{equation}\label{eq:KGM}
\mathcal{K}_L = \mathcal{G}_L + \mathcal{M}_L \left( I - \mathbf{T}_L^f \mathcal{G}_L \right),
\end{equation}
where
\begin{align}
\mathcal{G}_L &= \mathcal{T}_L \left( \Phi_1, B_1, C_1, D_1 \right), \label{eq:Gstar}\\
\mathcal{M}_L & = \mathcal{T}_L \left( \Phi_1- K_r C_2, K_r, C_1, 0 \right). \label{eq:Mstar}
\end{align}
By substituting (\ref{eq:KGM}) into (\ref{eq:festkL_inv}), the extended form (\ref{eq:KGM}) can be intuitively explaned:
\begin{enumerate}[(\romannumeral1)]
  \item $\mathcal{G}_L \mathbf{r}_{k,L}$ is the stacked fault estimates from the open-loop inverse (\ref{eq:open_loop_fest});
  \item $\mathbf{T}_L^f \mathcal{G}_L \mathbf{r}_{k,L}$ is the stacked reconstructed residuals generated by (\ref{eq:open_loop_fest_dyn}) and (\ref{eq:res_reconstruct});
  \item $\mathcal{M}_L$ represents the feedback dynamics from the \\stacked residual reconstruction errors $\mathbf{r}_{k,L} - \mathbf{T}_L^f \mathcal{G}_L \mathbf{r}_{k,L}$ to the stacked fault estimates.
\end{enumerate}

By substituting (\ref{eq:resL_compute}) and (\ref{eq:KGM}) into (\ref{eq:festkL_inv}), we obtain the following extended form of the fault estimation filter (\ref{eq:sfestfilter_reduced}):
\begin{subequations}
\begin{align}
{{\mathbf{\hat {{f}}}}_{k,L}} &= \mathcal{O}_L \left( \Phi_2, C_1 \right) \cdot \hat x(k_0) +
\left( \mathcal{R}_L + \mathcal{M}_L \mathcal{Q}_L \right)
\cdot \mathbf{z}_{k,L}, \label{eq:festfilter_extended} \\
&= \mathcal{O}_L \left( \Phi_2, -C_1 \right) \cdot \tilde x(k_0) + \mathbf{f}_{k,L}
+ \mathcal{K}_L \mathbf{e}_{k,L} \label{eq:festfilter_extended_dyn}
\end{align}
\end{subequations}
with
\begin{equation}\label{eq:RL_QL}
\begin{aligned}
\mathcal{R}_L &= \mathcal{G}_L \cdot \mathbf{T}_L^z = \mathcal{T}_L \left( \Phi_1, [\begin{matrix} B_f & K_f \end{matrix}], C_1,
                 [\begin{matrix} D_{f,1} & D_1 \end{matrix}] \right), \\
\mathcal{Q}_L &= \left( I - \mathbf{T}_{L}^f \mathcal{G}_L \right) \cdot \mathbf{T}_L^z
= \mathbf{T}_L^z - \mathbf{T}_{L}^f \mathcal{R}_L
\\ &= \mathcal{T}_L \left( \Phi_1, [\begin{matrix} B_f & K_f \end{matrix}], -C_2,
                 [\begin{matrix} -D_{f,2} & G_{f,2} \end{matrix}] \right).
\end{aligned}
\end{equation}
The derivation of $\mathcal{R}_L$ in (\ref{eq:RL_QL}) is obtained by regarding $\mathcal{G}_L \cdot \mathbf{T}_L^z$ as cascading the system $( \Phi, [\begin{array}{cc} \tilde B & K \end{array}], -C, [\begin{array}{cc} -D& I \end{array}] )$ related to $\mathbf{T}_L^z$ and the system $\left( \Phi_1, B_1, C_1, D_1 \right)$ related to $\mathcal{G}_L$. Similar methods apply to the derivation of $\mathcal{Q}_L$ in (\ref{eq:RL_QL}).

It can be seen from (\ref{eq:festfilter_extended_dyn}) that $\mathbf{\hat f}_{k,L}$ is a biased estimate of $\mathbf{f}_{k,L}$ due to the presence of unknown initial state $\tilde x(k_0)$. However, it follows from (\ref{eq:festfilter_extended_dyn}) and the definition of $\mathcal{O}_L \left( \Phi_2, -C_1 \right)$ in (\ref{eq:OL_TLmarkov}) that
\begin{equation*}
\begin{aligned}
\mathbb{E} \left( \hat f(k) - f(k) \right) = -C_1 \Phi_2^{L-1} \tilde x (k_0),
\end{aligned}
\end{equation*}
where $\hat f(k)$ and $f(k)$ are the last $n_f$ entries of $\mathbf{\hat f}_{k,L}$ and $\mathbf{f}_{k,L}$, respectively.
The above equation shows that $\hat f(k)$, extracted from $\mathbf{\hat f}_{k,L}$ in (\ref{eq:festfilter_extended}), gives asymptotically unbiased fault estimation as $L$ goes to infinity if $\Phi_2$ is stabilized given the condition in Theorem \ref{thm:stabilizability}.


\subsection{Markov-parameter based design}\label{sect:alg}
In order to avoid propagating model reduction errors that originate from the identification of the state-space plant model into the designed filter, we now directly construct the sensor fault estimation filter (\ref{eq:sfestfilter_reduced}) from data by utilizing its extended form (\ref{eq:festfilter_extended}).
The basic idea follows four steps:
\begin{enumerate}[(i)]
  \item Identify the Markov parameters $\Xi$ in (\ref{eq:Xi_Hi}) using the plant I/O data \citep{Chiuso2007, Chiuso2007a};
  \item Compute the Markov parameters of the system
        \begin{equation}\label{eq:IDfilter}
            \left( \Phi_1, [\begin{matrix} B_f & K_f \end{matrix}],
            \left[ \begin{matrix} C_1 \\ -C_2 \end{matrix} \right],
            \left[ \begin{matrix} D_{f,1} & D_1 \\ -D_{f,2} & G_{f,2} \end{matrix} \right] \right)
        \end{equation}
        which combines the dynamics related to $\mathcal{R}_L$ and $\mathcal{Q}_L$ in (\ref{eq:RL_QL});
  \item Find a state-space realization of the system (\ref{eq:IDfilter});
  \item Find a stabilizing gain $L_2$ for $\left( \Phi_1, C_2 \right)$, and then construct the fault estimation filter (\ref{eq:sfestfilter_reduced}) with the identified system matrices in (\ref{eq:IDfilter}).
\end{enumerate}

In Step (\romannumeral2), it should be noted that $\Xi$ in (\ref{eq:Xi_Hi}) includes only the Markov parameters $\{ H_i^u \}$ and $\{ H_i^y \}$ related to system inputs and outputs. According to (\ref{eq:fault_model_predictor}) and (\ref{eq:markov_param}), the Markov parameters $\{ H_i^f \}$ related to the  $j^{\mathrm{th}}$ sensor faults and $\{ H_i^z \}$ related to $\mathbf{T}_L^z$ in (\ref{eq:TLz}) can be obtained as
\begin{equation}\label{eq:Hiz}
  H_i^f = \left\{ \begin{array}{ll}
                   I^{[j]} & i=0 \\
                   -\left( H_i^y \right)^{[j]} & i>0
                 \end{array}
   \right. ,\;
H_i^z = \left\{ \begin{array}{ll}
                  \left[ \begin{matrix} -H_0^u & I \end{matrix} \right] & i=0 \\
                  \left[ \begin{matrix} -H_i^u & -H_i^y \end{matrix} \right] & i>0
                \end{array}
 \right. .
\end{equation}

Let $\{G_i\}$, $\{ R_i \}$ and $\{ Q_i \}$ denote the Markov parameters that construct the block-Toeplitz matrices
\begin{equation*}\label{eq:RL_QL_Markov}
\mathcal{\mathcal{G}}_L = \mathcal{T}_L \left( \{ G_i \} \right),\; \mathcal{R}_L = \mathcal{T}_L \left( \{ R_i \} \right), \; \mathcal{Q}_L = \mathcal{T}_L \left( \{ Q_i \} \right)
\end{equation*}
with the definition of $\mathcal{T}_L$ in (\ref{eq:OL_TLmarkov}).
In order to ensure $\mathcal{G}_L \mathbf{T}_L^f = I$, the Markov parameters $\{G_i\}$ can be computed as
\begin{equation}\label{eq:Gi}
\left\{
\begin{array}{l}
G_0 = \left( H_0^f \right)^-, \\
G_i = - \sum_{j=1}^i G_{i-j} H_j^f G_0, \; 1 \le i \le L-1.
\end{array}
\right.
\end{equation}

According to the definition of $\mathcal{R}_L$ in (\ref{eq:RL_QL}), its Markov parameters $\{ R_i \}$ can be computed as the convolution of $\{ G_i \}$ in (\ref{eq:Gi}) and $\{ H_i^z \}$ in (\ref{eq:Hiz}):
\begin{equation*}
R_i = \sum_{j=0}^i  G_{i-j} H_{j}^z \;\; i=0, \cdots, L-1.
\end{equation*}
Similarly, the Markov parameters $\{ Q_i \}$ of $\mathcal{Q}_L$ in (\ref{eq:RL_QL}) can be computed as
\begin{equation*}
Q_i = H_i^z - \sum_{j=0}^i  H_{i-j}^f R_{j} \;\; i=0, \cdots, L-1.
\end{equation*}
Since the system (\ref{eq:IDfilter}) combines the dynamics of $\mathcal{R}_L$ and $\mathcal{Q}_L$ in (\ref{eq:RL_QL}), the Markov parameters of the system (\ref{eq:IDfilter}) are $\{ W_i \}$ with
$W_i = \left[ \begin{smallmatrix} R_i \\ Q_i \end{smallmatrix} \right]$.

In Step (\romannumeral3), using the Markov parameters $\{ W_i \}$ of the system (\ref{eq:IDfilter}), it is straightforward to obtain
\begin{equation}\label{eq:DG}
\left[ \begin{matrix} \hat D_{f,1} & \hat D_1 \\ - \hat D_{f,2} & \hat G_{f,2} \end{matrix} \right] = W_0 =
\left[ \begin{matrix} R_0 \\ Q_0  \end{matrix} \right],
\end{equation}
and formulate the block Hankel matrix
\begin{equation}\label{eq:HRL}
\mathcal{H}_{W} = \left[
\begin{smallmatrix}
W_1 & W_2 & \cdots & W_m \\
W_2 & W_3 & \cdots & W_{m+1} \\
\vdots & \vdots & \ddots & \vdots \\
W_l & W_{l+1} & \cdots & W_{l+m-1}
\end{smallmatrix}
\right],
\end{equation}
which corresponds to the system (\ref{eq:IDfilter}).
Then, compute the singular value decomposition (SVD) of $\mathcal{H}_{W}$, i.e.,
\begin{equation*}\label{eq:HRsvd}
\mathcal{H}_{W} = \left[ \begin{matrix} U_W & U_W^{\bot} \end{matrix} \right]
 \left[ \begin{matrix}
                \Sigma_W & 0 \\
                0 & \Sigma_W^{\bot}
             \end{matrix} \right]
 \left[ \begin{matrix} V_W^\mathrm{T} \\ \left( V_W^{\bot} \right)^\mathrm{T} \end{matrix} \right].
\end{equation*}
In this above equation, the nonsingular and diagonal matrix $\Sigma_W$ consists of the $\hat n$ largest singular values of the block Hankel matrix $\mathcal{H}_W$, where $\hat n$ is actually the selected order of the fault estimation filter (\ref{eq:sfestfilter_reduced}). The order $\hat n$ can be chosen by examining the gap among the singular values of $\mathcal{H}_W$, as in subspace identification methods \citep{Chiuso2007}. Now, we can write
\begin{equation}\label{eq:HR_HQ_app}
\mathcal{\hat H}_{W} = U_W \Sigma_W V_W^\mathrm{T}.
\end{equation}

From (\ref{eq:HR_HQ_app}), the estimated controllability and observability matrices can be constructed as \citep{Chiuso2007}
\begin{equation}\label{eq:Ctrb_Obsv}
\mathcal{\hat C}_W = \Sigma_W^{\frac {1}{2}} V_W^\mathrm{T}, \;
\mathcal{\hat O}_W = U_W \Sigma_W^{\frac {1}{2}}. \;
\end{equation}
Then the state-space matrices of the system (\ref{eq:IDfilter}) are computed as below:
\begin{align}
[\begin{matrix} \hat B_f & \hat K_f \end{matrix}] &= \text{the first}\;{n_u+n_y}\;\text{columns of}\;\mathcal{\hat C}_W, \label{eq:BfKf}\\
\left[\begin{smallmatrix} \hat C_{1} \\ - \hat C_2 \end{smallmatrix} \right] &=
\text{the first}\;{n_f+n_y}\;\text{rows of}\;\mathcal{\hat O}_W, \label{eq:C1C2} \\
\hat \Phi_1 &= \mathcal{\hat C}_{W,2} \mathcal{\hat C}_{W,1}^\mathrm{T} \left( \mathcal{\hat C}_{W,1} \mathcal{\hat C}_{W,1}^\mathrm{T} \right)^{-1}, \label{eq:Phiest}
\end{align}
where $\mathcal{\hat C}_{W,1}$ and $\mathcal{\hat C}_{W,2}$ are the matrices consisting of the first and, respectively, the last $n_u \left( m-1 \right)$ columns of $\mathcal{\hat C}_{W}$.

Finally, we follow Step (\romannumeral4) by using the estimated state-space matrices in (\ref{eq:DG}) and (\ref{eq:BfKf})-(\ref{eq:Phiest}).


\subsection{Comparisons and discussion}\label{Sect:comp_disc}

Fault reconstruction based on dynamic latent variable models has been widely adopted in statistical process monitoring \citep{Qin2012}.
In these types of methods, fault estimates are obtained by minimizing the squared reconstructed prediction error in the residual subspace of a latent variable model. By analyzing its close link with subspace identification, \cite{QinLi2001} showed that the dimension of the residual subspace is determined by the left null space of the observability matrix, thus limiting fault reconstructability for dynamic latent variable models.

Although the stabilizability condition in Theorem \ref{thm:stabilizability} is derived by using the predictor model (\ref{eq:predictor}), it can still be applied to the data-driven filter design without requiring the state-space model, because the invariant zeros of the underlying fault subsystem $( \Phi, \tilde E, C, G )$ can be checked by using the identified Markov parameters $\{ H_i^f \}$ \citep{Yeung1993, Fled2010}. In this sense, the stabilizability condition of our proposed data-driven filter design is the same as in model-based filter design.

The additional design freedom for stabilization is not possible in the data-driven fault estimation filter design proposed by \cite{Dong2012c}, because it considered only the open-loop left inverse (\ref{eq:open_loop_fest}) and missed the feedback part in the inverse dynamics.
For the same reason, the data-driven filter in \cite{Dong2012c} would be unstable in some situations, such as sensor faults of an unstable open-loop plant. In contrast, our data-driven design is based on the closed-loop left inverse (\ref{eq:sysinv_closed_loop}) which guarantees the stability and unbiasedness of the constructed fault estimation filter under the condition given by Theorem \ref{thm:stabilizability}.

The above discussion explains why sensor faults of an unstable open-loop plant cannot be tackled by applying the data-driven method in \cite{Dong2012c} to the open-loop plant.
It is worth noting that this difficulty cannot be solved by simply applying the same method to the stabilized closed-loop system. The reason is that the sensor faults affect not only output equations but also the closed-loop dynamics, hence the first equation in (\ref{eq:Hiz}) is no longer valid for the Markov parameters $\{ H_i^f \}$ of the closed-loop fault subsystem. In fact, with only fault-free I/O data as stated in Assumption \ref{ass:data}, there is no simple way to derive $\{ H_i^f \}$ from the identified Markov parameters $\{ H_i^u, H_i^y \}$ of a closed-loop system \citep{Wan2012}. In order to use the closed-loop system for data-driven sensor fault estimation, Section \uppercase\expandafter{\romannumeral5}.B of \cite{Dong2012c} proposed to use a special control law such that the sensor faults did not affect the closed-loop dynamics, which is not always possible in practice.

It is also interesting to investigate the computational complexity of our proposed
fault estimation filter (\ref{eq:sfestfilter_reduced}), which turns out to be $O \left( (n_f + n_x) (n_x + n_u + n_y) \right)$ at each time instant.
In order to put this in perspective, we compare it with our recently proposed moving horizon fault estimator in \cite{WanKevicVerh2014},
where choosing a certain horizon length $L$, the computations at each time instant involve multiplying a matrix of size $n_f \times L(n_u + n_y)$ with $\mathbf{z}_{k,L}$ defined in (\ref{eq:ukL}).
This leads to a computational complexity of $O(L n_f (n_u+n_y))$. Therefore, by directly identifying the fault estimation filter (\ref{eq:sfestfilter_reduced}), our proposed new method would be more efficient in
terms of computational complexity, since $L$ needs to be sufficiently large in order to achieve asymptotic unbiasedness in \cite{WanKevicVerh2014}.

We show in Appendix~\ref{app:eqF} that our recently proposed moving horizon fault estimator in \cite{WanKevicVerh2014} can be equivalently rewritten as
\begin{equation}\label{eq:F}
\mathbf{\hat f}_{k,L}^{'} = \left[ \mathcal{G}_L^{'} + \mathcal{M}_L^{'} \left( I - \mathbf{T}_L^f \mathcal{G}_L^{'} \right) \right] \cdot \mathbf{r}_{k,L}
\end{equation}
which is in the same form of (\ref{eq:festkL_inv}) and (\ref{eq:KGM}) except that $\mathcal{G}_L$ and $\mathcal{M}_L$ in (\ref{eq:KGM}) have the block-Toeplitz structure defined in (\ref{eq:Mstar}) while $\mathcal{G}_L^{'}$ and $\mathcal{M}_L^{'}$ in (\ref{eq:F}) are dense matrices according to (\ref{eq:G})-(\ref{eq:Delta}). Compared to the moving horizon fault estimator in \cite{WanKevicVerh2014}, the block-Toeplitz structure of $\mathcal{G}_L$ and $\mathcal{M}_L$ in our proposed new design affects two aspects:
\begin{enumerate}[(i)]
  \item It has been proved in \cite{Wan2014} that with dense matrices $\mathcal{G}_L^{'}$ and $\mathcal{M}_L^{'}$ in (\ref{eq:F}), the LS moving horizon fault estimation used in \cite{WanKevicVerh2014} gives minimal variance over all linear unbiased estimators. In contrast, the block-Toeplitz structure of $\mathcal{G}_L$ and $\mathcal{M}_L$ in our proposed new method leads to less design freedom and minor performance loss.
  \item In order to ensure asymptotically unbiased estimation, the moving horizon fault estimator in \cite{WanKevicVerh2014} requires that the fault subsystem has no unstable transmission zeros, while a more strict condition is needed for our proposed new design, i.e., not only the transmission zeros but all the invariant zeros of the fault subsystem have to be stable.
\end{enumerate}

\section{Simulation studies}
Consider the linearized continuous-time VTOL (vertical takeoff and landing) aircraft model used in \cite{Dong2012c}.
The model has four states, namely horizontal velocity, vertical velocity, pitch rate, and pitch angle. The two inputs are collective pitch control and longitudinal cyclic pitch control.
With a sampling rate of 0.5 seconds, the discrete-time model (\ref{eq:sys}) is obtained, with $D = 0$ and $F = I_4$.
The process and measurement noise, $w(k)$, $v(k)$, are zero mean white, respectively with a covariance of $Q_w = 10^{-4} \cdot I_4$ and $Q_v = 0.01 \cdot I_2$.
Since the open-loop plant is unstable, an empirical stabilizing output feedback controller is used, i.e.,
\begin{equation}\label{eq:controller}
u(k) = - \left[ \begin{smallmatrix}
                  0 & 0 & -0.5 & 0 \\
                  0 & 0 & -0.1 & -0.1
                \end{smallmatrix}
 \right]\cdot y(k) + \eta(k),
\end{equation}
where $\eta(k)$ is the reference signal. All the parameters of the plant and the controller are the same as those in \cite{Dong2012c}.

In the identification experiment, the reference signal $\eta(k)$ is zero-mean white noise with the covariance of $\mathrm{diag}\left( 1, 1 \right)$, which ensures persistent excitation.
We collect $N=1000$ data samples from the identification experiment. In the identification algorithm, the past horizon is selected as $p=100$.

The simulated sensor fault signals affect the first two sensors, i.e., $E=0_{4 \times 2}$,
$G = \left[ \begin{smallmatrix} 1 0 0 0 \\ 0 1 0 0 \end{smallmatrix} \right]^\mathrm{T}$,
$$ f(k) = \left\{ \begin{array}{ll}
   \left[ \begin{array}{cc}
            0 & 0
           \end{array}
   \right]^\mathrm{T}, & 0 \leq k \leq 50, \\
   \left[ \begin{array}{cc}
            \mathrm{sin}\left( 0.1 \pi k \right) & 1
           \end{array}
   \right]^\mathrm{T}, & k > 50.
   \end{array}
   \right. $$
When simulating sensor faults, the reference signal $\eta (k)$ in the control law (\ref{eq:controller}) is set to be zero.

We will compare the following methods for fault estimation filter design:
\begin{itemize}
    \item Alg0: design based on the accurate predictor model (\ref{eq:predictor});
    \item Alg1: design based on the state-space model of the predictor (\ref{eq:predictor}) identified from data;
    \item Alg2: our proposed new method in Section \ref{sect:alg};
    \item Alg3: our recently proposed moving horizon fault estimator constructed from the predictor Markov parameters $\Xi$ identified from data \citep{WanKevicVerh2014}.
\end{itemize}
In the state-space realization from Hankel matrices, the orders of the plant model in Alg1 and the filter in Alg2 are both selected as 4, the same as the underlying plant. For Alg2 and Alg3, the order of the Markov parameters of the system (\ref{eq:IDfilter}) is $L = 100$, and the number of block rows and columns of the block Hankel matrix $\mathcal{H}_W$ in (\ref{eq:HRL}) is $l = m = 20$. The poles of the first three fault estimation filters are placed at the same location, i.e., $[ 0.948, 0.532, 0.225, 0.141 ]^{\mathrm{T}}$.

The method in \citep{Dong2012c} cannot be directly applied to the above sensor fault scenario due to the reasons explained in Section \ref{Sect:comp_disc}, thus it is not implemented here. In contrast, all the three fault estimation filters above are based on the closed-loop left inverse (\ref{eq:sysinv_closed_loop}), and their stability is guaranteed since the condition in Theorem \ref{thm:stabilizability} is satisfied in this simulated scenario. 

\begin{figure}[h]
\centering
\includegraphics[width=7cm]{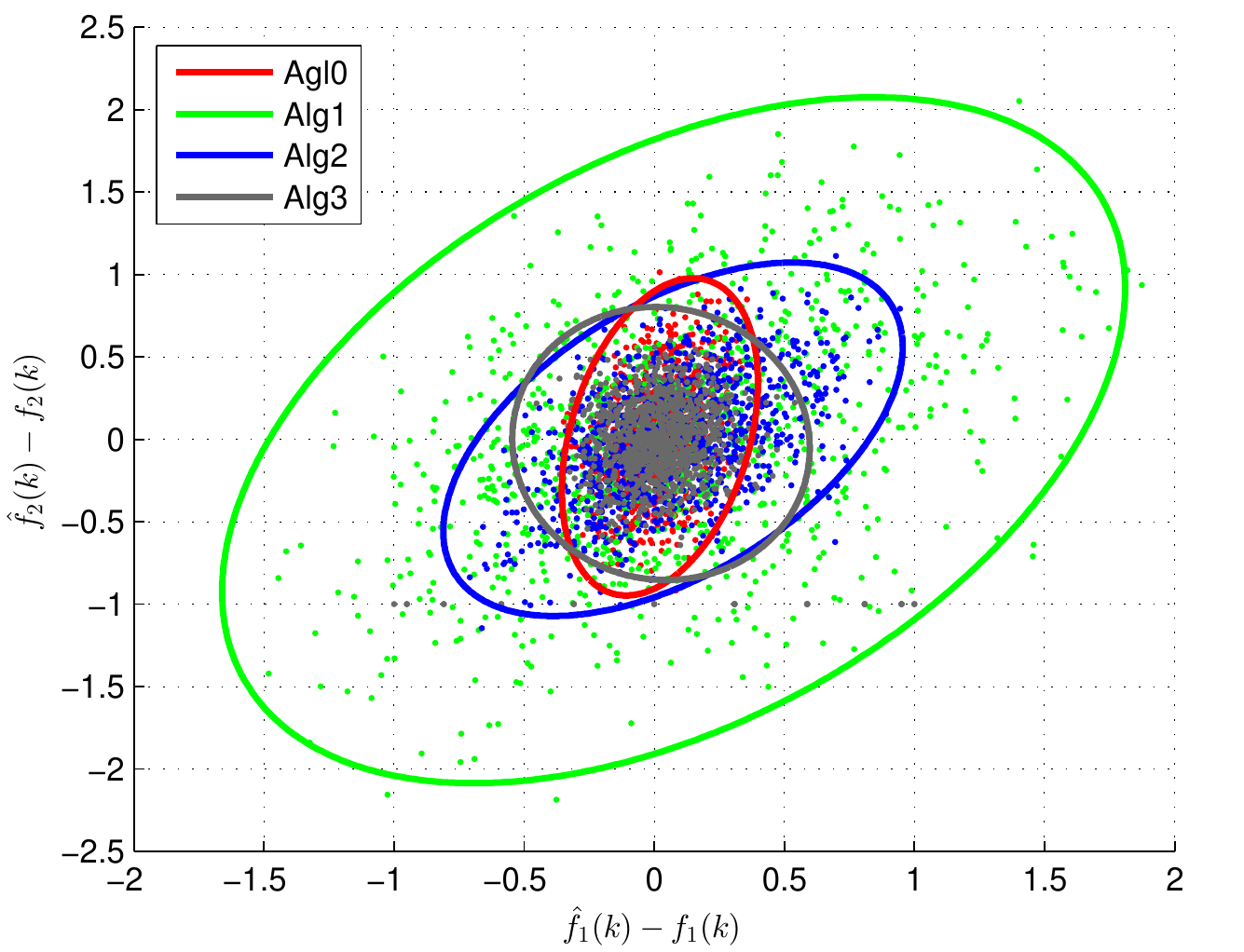}
\caption{Distribution of fault estimation errors. Dots: 1000 estimation errors by different fault estimation methods. Ellipses: the $3\sigma$-contour of the approximated two-dimensional Gaussian distribution of the 1000 estimation errors, i.e., the contour at $[ \hat f(k) - f(k) ]^\mathrm{T} {\text{cov}^{-1} ( \hat f(k) )}
[ \hat f(k) - f(k) ] = 3$.}
\label{fig:result}
\end{figure}

The distributions of fault estimation errors are shown in Figure \ref{fig:result}. Because of the noisy identification data and the model reduction errors, the three data-driven designs, Alg1, Alg2, and Alg3, all give larger estimation error covariances than Alg0 based on the accurate plant model. Figure \ref{fig:result} also clearly shows that our proposed Alg2 achieves better estimation performance than Alg1. This is because Alg2 is not subject to model reduction errors before realizing the state-space matrices in (\ref{eq:IDfilter}) as part of the fault estimation filter (\ref{eq:sfestfilter_reduced}), while model reduction errors are introduced instead by Alg1 in the identified plant model, and propagated into larger uncertainties of the fault estimation filter. As explained at the end of Section \ref{Sect:comp_disc}, Alg2 is much faster than Alg3 at the cost of minor performance loss.

\section{Conclusions}
A novel direct data-driven design method has been proposed for sensor fault estimation filters by parameterizing the system-inversion-based fault estimation filter with Markov parameters.
The proposed approach simplifies the design procedure by omitting the step of identifying the state-space plant model, and improves estimation performance by avoiding the propagation of some model reduction errors into the designed filter. Moreover, it allows additional design freedom to stabilize the designed filter under the same stabilizability condition as model-based system inversion, thus can be applied to sensor faults in an unstable plants. Detailed analysis has been given to explain why sensor faults in an unstable plant cannot be tackled by applying existing state-of-the-art data-driven methods 
with predictor Markov parameters of either the unstable open-loop plant or the stabilized closed-loop system.
A numerical simulation example illustrates the effectiveness of our method applied to sensor faults of an unstable aircraft system, and the advantage of the direct data-driven design.
Future work will focus on how to enhance robustness against both identification errors of Markov parameters and model reduction errors in realizing the state-space form of the filter.


\bibliography{fest_filter_ID_arXiv}             

\appendix
\section{Proof of Theorem \ref{thm:stabilizability}}\label{app:proof_stabilizability}    
In order to prove $\left( \Phi_1, C_2 \right)$ is stabilizable, we need to show that $\left( \Phi_1, C_2 \right)$ has no unstable unobservable modes, i.e.,
\begin{equation}\label{eq:unobsv_mode}
\text{rank} \left( \left[ \begin{array}{c}
                          \Phi_1 - \lambda I \\
                          C_2
                        \end{array}
 \right] \right) = n \;\text{for}\; \left| \lambda \right| \ge 1.
\end{equation}

By following (\ref{eq:Phi1_B1}) and (\ref{eq:C2_D2}), it can be derived that
\begin{align}
\left[ \begin{array}{cc}
         \Phi_1 - \lambda I & \tilde E \\
         C_2 & G
       \end{array}
 \right]
= \left[ \begin{array}{cc}
         \Phi - \lambda I & \tilde E \\
         C & G
       \end{array}
 \right] \cdot
 \left[ \begin{array}{cc}
         I & 0 \\
         -G^- C & I
       \end{array}
 \right].
\end{align}
With Assumption \ref{ass:fault_rank}, if $\left( \Phi, \tilde E, C, G \right)$ has no unstable invariant zeros, it follows that
\begin{equation*}
\begin{aligned}
&\text{rank} \left( \left[ \begin{array}{cc}
         \Phi_1 - \lambda I & \tilde E \\
         C_2 & G
       \end{array}
 \right] \right) =
\text{rank} \left( \left[ \begin{array}{cc}
         \Phi - \lambda I & \tilde E \\
         C & G
       \end{array}
 \right] \right)
= n + n_f
\end{aligned}
\end{equation*}
for $\left| \lambda \right| \ge 1$, which implies (\ref{eq:unobsv_mode}).

\section{Proof of (\ref{eq:F})}\label{app:eqF}    
With the definitions
\begin{equation}\label{eq:Psi_fx}
\mathbf{O}_L = \mathcal{O}_L (\Phi, C),\;
\Psi_L = \left[  \begin{matrix}
  \mathbf{O}_L & \mathbf{T}_L^f
\end{matrix}  \right],\;
\mathbf{f}_{k,L}^x = \left[ \begin{matrix}
  \tilde x(k_0) \\ \mathbf{f}_{k,L}
\end{matrix} \right],
\end{equation}
the LS fault estimation problem
\begin{equation}\label{eq:LS_prob}
\mathop {\min }\limits_{{{\mathbf{{{f}}}}_{k,L}^x}} \;\left\| {{{\bf{r}}_{k,L}} - \Psi_{L}
\mathbf{{{f}}}_{k,L}^x } \right\|_{2}^2
\end{equation}
can be formulated based on the second equation of (\ref{eq:resL_compute}) \citep{Wan2014}.
The LS problem (\ref{eq:LS_prob}) has non-unique solutions because $\Psi_{L}$ may not have full column rank. By using the generalized inverse defined in (\ref{eq:ginv}), one solution to the problem (\ref{eq:LS_prob}) is
\begin{equation}\label{eq:fxkL_hat}
{{\mathbf{\hat {{f}}}}_{k,L}^x} = \left( \Psi_{L}^\mathrm{T} \Psi_{L} \right)^{(1)}
\Psi_{L}^\mathrm{T} {{\bf{r}}_{k,L}}.
\end{equation}
According to the definition of $\Psi_{L}$ in (\ref{eq:Psi_fx}) and Schur complements \citep{Kailath2000}, the estimate of $\mathbf{f}_{k, L}$, i.e., (\ref{eq:F}), can be extracted from (\ref{eq:fxkL_hat}), with
\begin{align}
\mathcal{G}_L^{'} &= \left( \left(\mathbf{T}_{L}^f\right)^\mathrm{T} \mathbf{T}_{L}^f \right)^{-1} \left( \mathbf{T}_{L}^f \right)^\mathrm{T}, \label{eq:G}\\
\mathcal{M}_L^{'} &= \mathcal{G}_L^{'} \mathbf{O}_L \Delta^{(1)} \mathbf{O}_L^\mathrm{T} \label{eq:M}, \\
\Delta &= \mathbf{O}_L^\mathrm{T} \mathbf{O}_L - \mathbf{O}_L^\mathrm{T}
\mathbf{T}_{L}^f \mathcal{G}_L^{'} \mathbf{O}_L. \label{eq:Delta}
\end{align}
Although $\mathbf{\hat f}_{k, L}^{'}$ in (\ref{eq:F}) may be a biased estimate of $\mathbf{f}_{k,L}$ due to column rank deficiency of $\Psi_{L}$, its last $n_f$ entries give an asymptotically unbiased estimate of $f(k)$ as $L$ goes to infinity \citep{Wan2014}.

Note that with the innovation covariance matrix estimated from data \citep{Chiuso2007, Chiuso2007a}, a weighted LS problem can be formulated for fault estimation \citep{Wan2014}. For the sake of simplicity, an ordinary LS problem (\ref{eq:LS_prob}) is used here without loss of generality.

\end{document}